\documentclass{article} 
\usepackage{spconf}
\usepackage[utf8]{inputenc}
\usepackage[squaren,Gray]{SIunits}
\usepackage{cite}
\usepackage{amssymb,upgreek,amsmath,bm,bbm}
\usepackage{amsthm, algorithm2e}
\usepackage{graphicx, epstopdf, tikz,pgfplots}
\usepgfplotslibrary{groupplots}
\usetikzlibrary{intersections,calc,arrows,matrix,spy,math,calc,fit}
\pgfplotsset{compat = 1.16}

\usepackage[justification=centering]{subfig}
\usepackage[export]{adjustbox}
\usepackage{subfig}
\usepackage{datatool}
\usepackage{array}
\graphicspath{{figs/}}
\usepackage{mystyle}
\usepackage[page]{appendix}
\usepackage{hyperref}

\usepackage{titlesec}
\titleformat*{\paragraph}{\itshape\mdseries}

\title{Optimal-transport-based metric for SMLM}
\name{Quentin Denoyelle$^\dagger${}$^\ddagger$,\thanks{
Q.D., T-a.P. and M.U. were supported by the European Research Council (ERC) under the European Union’s Horizon 2020 research and innovation programme (Grant Agreement No. 692726 GlobalBioIm).
}
Thanh-an Pham$^\dagger$,
Pol {del Aguila Pla}$^\diamond${}$^\dagger$, Daniel Sage$^\dagger$, and Michael Unser$^\dagger$}
\address{$^\dagger$ Biomedical Imaging Group, \'Ecole polytechnique fédérale de Lausanne, Lausanne, Switzerland\\
$^\diamond$ CIBM Center for Biomedical Imaging, Switzerland\\
$^\ddagger$ MAP5, Université de Paris, Paris, France.
		 }

\begin{document}

\maketitle
\ninept 

\begin{abstract}
We propose the use of Flat Metric to assess the performance of reconstruction methods for single-molecule localization microscopy (SMLM)
in scenarios where the ground-truth is available. Flat Metric is intimately
related to the concept of optimal transport between measures of different mass, providing solid 
mathematical foundations for SMLM evaluation and integrating both localization and detection 
performance. In this paper, we provide the foundations of Flat Metric and validate this measure by applying it to controlled synthetic examples and to data from the SMLM 2016 Challenge.
\end{abstract}
%

\section{Introduction}
\label{sec:intro}

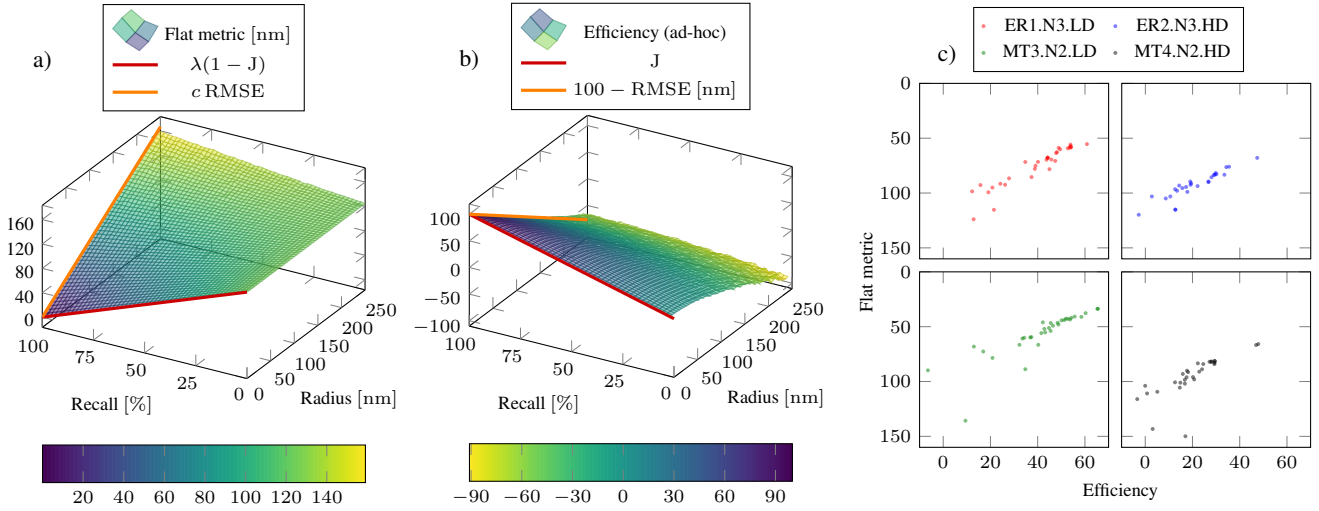
\begin{figure*}
    \centering
    	\begin{tabular}{m{.28\textwidth}m{.31\textwidth}m{.34\textwidth}}

\begin{tikzpicture}
        \pgfplotsset{colormap/viridis}
            \begin{axis}[
            width=.33\textwidth,
            scaled z ticks=manual:{}{},
            scaled x ticks=manual:{}{},
            view={-60}{40},
	    ticklabel style = {font=\scriptsize},
            xtick={0,0.5E-7,1E-7,1.5E-7,2E-7,2.5E-7},
            xticklabels={{$0$},{$50$},{$100$},{$150$},{$200$},{$250$}},
            ytick={0,25,50,75,100},
            ztick={0,0.4E-7,0.8E-7,1.2E-7,1.6E-7},
            zticklabels={0, 40, 80, 120, 160},
            xlabel={Radius~$[\mathrm{nm}]$},
            ylabel={Recall~$[\%]$},
            label style={font=\scriptsize},
            xlabel style={xshift=-7pt,yshift=5pt,},
            ylabel style={yshift=5pt,},
            colorbar horizontal,
            colorbar style={
		scaled x ticks=manual:{}{},
		xtick={0.2E-7,0.4E-7,0.6E-7,0.8E-7,1E-7,1.2E-7,1.4E-7},
                 xticklabels={{$20$},{$40$},{$60$},{$80$},{$100$},{$120$}, {$140$}},
                 ticklabel style = {font=\scriptsize},
            },
            legend style={font=\scriptsize,anchor=center, at={(0.5,1.225),},},
            ]
            \addplot3 [
            	surf,
		opacity=0.5,
            ] table {figs/OTresultfiles/UOT_vs_recall_and_radius_result-6.40E-06-1.25E-07-100-50-51-51.txt};
                \addlegendentry{Flat metric~$[\mathrm{n}\metre]$}
            \addplot3 [
            	red!80!black,
		very thick,
	    ] coordinates {(0,100,0) (0,0,125E-9)};
                \addlegendentry{$\lambda (1-\mathrm{J})$}
            \addplot3 [
            	orange, 
		very thick
	    ] coordinates{(0,100,0) (2.5E-7,100,1.667E-7)};
                \addlegendentry{$c\, \mathrm{RMSE}$}
            \end{axis}
            \node at (0,4.225) {a)};
\end{tikzpicture}&

\begin{tikzpicture}
            \begin{axis}[
            width=.33\textwidth,
            scaled z ticks=manual:{}{},
            scaled x ticks=manual:{}{},
            view={-60}{40},
	    ticklabel style = {font=\scriptsize},
	    xtick={0,0.5E-7,1E-7,1.5E-7,2E-7,2.5E-7},
	    xticklabels={{$0$},{$50$},{$100$},{$150$},{$200$},{$250$}},
	    ytick={0,25,50,75,100},
	    ztick={-100,-50,0,50,100},
	    xlabel={Radius~$[\mathrm{nm}]$},
            ylabel={Recall~$[\%]$},
            label style={font=\scriptsize},
            xlabel style={xshift=-7pt,yshift=5pt,},
            ylabel style={yshift=5pt,},
            colormap={reverse jet}{
                indices of colormap=(\pgfplotscolormaplastindexof{viridis}, ..., 0 of viridis)
            },
            colorbar horizontal,
            colorbar style={
                xtick={-90,-60,-30,0,30,60,90},
                ticklabel style = {font=\scriptsize},
            },
            legend style={font=\scriptsize,anchor=center, at={(0.5,1.225),},}
            ]
            \addplot3 [
	            surf,
	            opacity=0.5,
            ] table {figs/OTresultfiles/EFF_vs_recall_and_radius_result-6.40E-06-1.25E-07-100-50-51-51.txt};
                \addlegendentry{Efficiency (ad-hoc)}
            \addplot3 [
            	red!80!black,
		very thick,
	    ] coordinates {(0,100,100) (0,0,0)};
                \addlegendentry{$\mathrm{J}$}
            \addplot3 [
            	orange, 
		very thick
	    ] coordinates{(0,100,100) (2.5E-7,100,-76.78)};
                \addlegendentry{$100-\mathrm{RMSE}~[\mathrm{nm}]$}
            \end{axis}
            \node at (0,4.225) {b)};
\end{tikzpicture}&

 \newcommand{\marksize}{0.75pt}
   
\begin{tikzpicture}
	\begin{axis}[name = wtf1,
    		width = 0.33\textwidth,
   	 	height = 2.65cm,
    		ticks = none, draw=none, hide axis,
		legend columns=2,
    		legend style={
			at={(0.5,0.38)},
    			anchor=base,
    			column sep=0.5em,
    			legend style={font=\scriptsize,},
  		},
	]
    		\addlegendimage{only marks,mark=*,red,mark size=\marksize,opacity=0.5,
			draw opacity=0,};
    		\addlegendentry{ER1.N3.LD};
    		\addlegendimage{only marks,mark=*,blue,mark size=\marksize,opacity=0.5,
			draw opacity=0,};
    		\addlegendentry{ER2.N3.HD};
    		\addlegendimage{only marks,mark=*,green!50!black,mark size=\marksize,opacity=0.5,
			draw opacity=0,};
    		\addlegendentry{MT3.N2.LD}
    		\addlegendimage{only marks,mark=*,black,mark size=\marksize,opacity=0.5,
			draw opacity=0,};
    		\addlegendentry{MT4.N2.HD};
		\addplot [draw=none] {x};
    	\end{axis}
	\node at (0,0.38) {c)};
	\begin{groupplot}[
		group style={
			group name=wtf, 
			group size={2 by 2},
			vertical sep=5pt, 
			horizontal sep=5pt,
		},
		width=0.23\textwidth,
		height=0.22\textwidth,
  		x label style={at={(axis cs:75,185)},anchor = north},
   		y label style={at={(axis cs:-30,0)},anchor = base},
		label style={font=\scriptsize},
		xmin = -10, xmax=70,
    		ymin = 0, ymax=160,
		ticklabel style = {font=\scriptsize},
    	]
	\nextgroupplot[
		y dir=reverse,
		xticklabels={},
    		at = (wtf1.south),
    		anchor = north east,
    	]
		\addplot[
			only marks,
    			mark=*,
    			color=red,
    			mark size=\marksize,
			opacity=0.5,
			draw opacity=0,
		] table [x=efficiency, y=ot, col sep=comma] {figs/challenges/ER1.N3.LD.csv};
	\nextgroupplot[
		y dir=reverse,
		xticklabels={},
		yticklabels={},
	]
    		\addplot[
			only marks,
    			mark=*,
    			color=blue,
    			mark size=\marksize,
			opacity=0.5,
			draw opacity=0,
		] table [x=efficiency, y=ot, col sep=comma] {figs/challenges/ER2.N3.HD.csv};
	\nextgroupplot[
		y dir=reverse,
		ylabel={Flat metric},
    		xlabel={Efficiency},
	]
    		\addplot[
			only marks,
    			mark=*,
    			color=green!50!black,
    			mark size=\marksize,
			opacity=0.5,
			draw opacity=0,
		] table [x=efficiency, y=ot, col sep=comma] {figs/challenges/MT3.N2.LD.csv};
    	\nextgroupplot[
		y dir=reverse, 
		yticklabels={},
	]
    	\addplot[
		only marks,
    		mark=*,
    		color=black,
    		mark size=\marksize,
		opacity=0.5,
		draw opacity=0,
	] table [x=efficiency, y=ot, col sep=comma] {figs/challenges/MT4.N2.HD.csv};
    \end{groupplot}
\end{tikzpicture}
        \end{tabular}

        \vspace{-5pt}
            
        \caption{a) Flat Metric (low: good). b) Efficiency (low: bad). These metrics take into account a continuum in both localization and detection errors in SMLM.
        Locations (100) were uniformly drawn to create ground-truth points and modified to create artificial sets of detections with $100\%$ precision, recall ranging from $0\%$ to $100\%$, and localization errors uniformly sampled in a circle of a given radius.
        c) High degree of correlation between efficiency and Flat Metric on four datasets from the SMLM 2016 Challenge.
         \label{fig:surface}}
\end{figure*}

Single-molecule localization microscopy (SMLM) has boomed over the past decade, delivering on the promise of breaking the diffraction limit and giving access to otherwise unreachable cellular structures~\cite{sauer2017single,laketa2018microscopy}.
SMLM relies on computational methods that detect and accurately localize the few fluorescent emitters in each of many acquired frames, ultimately creating a superresolved image (up to $10$nm)~\cite{Sage2019}.
Therefore, it is crucial to have at one's disposal an objective evaluation of the recovery performance of available reconstruction algorithms.
The present paper studies this topic, under the hypothesis that a ground-truth reference for every captured frame is available. Metrics that do not require ground-truth information also exist~\cite{nieuwenhuizen2013measuring, banterle2013fourier, culley2018quantitative,descloux2019parameter}, even some using optimal transport concepts~\cite{Mazidi2020}. However, these are outside of the scope of this paper, and our proposal is completely new. Similarly, simpler optimal-transport-based metrics were used before in other point-source localization problems~\cite{AguilaPla2017,AguilaPla2017a}.

The localization of point sources is traditionally assessed using either detection metrics, such as precision, recall, and the Jaccard index; or localization metrics, such as the root-mean-square error (RMSE) or the root-mean-square minimum distance~(RMSMD)~\cite{sun2018root}. 
In the SMLM 2016 Challenge~\cite{Sage2019}, a large panel of metrics was computed for performance assessment.
The participating localization algorithms typically focused on one of two main key metrics: the Jaccard index~(J) or the root-mean-square error~(RMSE).
To encompass both, Sage \emph{et al.} proposed the efficiency, a metric born from the analysis of the empirical results in~\cite{Sage2019} and designed to evaluate the SMLM 2016 Challenge. It is computed as
\begin{equation}\label{eq:efficiency}
\mathrm{efficiency} = 100 - \sqrt{(100 - \mathrm{J})^2 + \alpha_\mathrm{eff}^2\mathrm{RMSE}^2}.
\end{equation}
The parameter~$\alpha_\mathrm{eff}$ was introduced to regulate the tradeoff between localization and detection.
It was set to~\mbox{$\alpha_\mathrm{eff} = 1 \nano\meter^{-1}$} for the two-dimensional (lateral) efficiency after analysis of the results for the best algorithms. 
With this empirical choice, an improvement of $1\nano\meter$ in RMSE is equivalent to a $1$\% improvement in~J.

In this paper, we propose to use~\emph{Flat Metric}, also known in the literature as the flat norm or the Kantor-Rubinstein norm~\cite{federer2014geometric,Peyre2019Computational,hanin1992kantorovich,lellmann2014imaging}
, to assess the recovery performance of algorithms for SMLM.
This metric has already been used to assess the recovery performance of point source signals~\cite{catala2019low}.
It can be related to optimal transport which is a well-studied field both on a theoretical~\cite{villani2008optimal,santambrogio2015optimal} and numerical~\cite{Peyre2019Computational} standpoint.
By using a valid metric on the space of Radon measures, in which detections and ground-truth data lie, we expose the natural connection between the localization-detection performance tradeoff and the radius of tolerance used to judge a detected location as correct or incorrect.
Furthermore, like other metrics introduced recently for SMLM~\cite{sun2018root}, Flat Metric does not require arbitrary pairing decisions between detected and ground-truth locations.
Nonetheless, in opposition to RMSMD, Flat Metric still resolves pairings implicitly, thus yielding interpretable and explainable assessments.

The paper is structured as follows: First, we introduce Flat Metric mathematically, 
expose its link with
unbalanced optimal transport and explain how to compute it numerically. Then, we illustrate its behavior on a simple example. Finally, we compare it to the efficiency~\eqref{eq:efficiency} on both synthetic data and the SMLM 2016 challenge data.


%

\section{Flat Metric for SMLM} \label{sec:math}

\subsection{Mathematical Definition}
\label{sec:mathdef}
\def\dim{D}

Without loss of generality, we assume that the ground-truth and detected locations are in $\Xx=[0,1]^\dim$, for $\dim\in\lbrace2,3\rbrace$.
We use the Euclidean distance $d(\V x,\V y)=\norm{\V x - \V y}_2$ to measure the distances between two points.
We denote by $\Radon$ the space of Radon measures defined on $\Xx$. Mathematically, $\Radon$ is the continuous dual of the space $\Cc(\Xx)$ of continuous functions on $\Xx$ endowed with the uniform norm $\normi{\cdot}$. The canonical norm on $\Radon$ is thus
\eq{
	\forall \mu\in\Radon, \quad \mnorm{\mu}\triangleq\underset{f\in\Cc(\Xx),\normi{f}\leq 1}{\sup} \int_\Xx f \mathrm{d}\mu,
}
and is known as the total-variation norm or $\Mm$ norm. The Banach space $\Radon$ contains point-source signals, referred to as the Dirac masses $\dirac{\V x} \triangleq \delta(\cdot-\V x)$ for $\V x\in\Xx$. This makes it particularly well-suited for SMLM because individual fluorescent emitters can be seen as Dirac masses, which suggests the representation of SMLM data as sums of Dirac masses.

The total-variation norm is not a good candidate metric for SMLM because, for all $\V x\neq\V y$, $\mnorm{\dirac{\V x} - \dirac{\V y}} = 2$. Instead, we build our metric from the flat norm on $\Radon$ given in Definition~\ref{def:flatnorm}.

\begin{definition}[Flat norm~\cite{hanin1992kantorovich}]\label{def:flatnorm}
The flat norm of a given $\mu\in\Radon$ is defined as
\begin{align}
	\norm{\mu} \triangleq \sup\left(\left\{\!\int_\Xx f \mathrm{d}\mu \!:\! f\in\Cc(\Xx),\normi{f}\leq \la, \Lip(f)\leq1\right\}\right)\!,
\end{align}
where $\Lip(f)$ is the Lipschitz constant of $f$. This definition induces a norm on $\Radon$.
\end{definition}

Using the flat norm to measure the difference between two Radon measures leads to Flat Metric.

\begin{definition}[Flat Metric]\label{def:flatmetric}
Flat Metric is defined for any two $\mu,\nu\in\Radon$ as
\begin{align}\label{eq:def-flatmetric-general}
	\KR{\mu}{\nu}\triangleq\norm{\mu - \nu}.
\end{align}
\end{definition}

Flat Metric is linked to unbalanced optimal transport~\cite{schmitzer2017framework,Peyre2019Computational}. This makes Flat Metric interpretable, which is key for its application to SMLM. 

\begin{proposition}[Interpretation of Flat Metric - \cite{schmitzer2017framework}, Prop. 2.26]\label{prop:uot}
For all $\mu,\nu\in\Radon$,
\begin{align}\label{eq:dual-primal-continuous}
	\KR{\mu}{\nu}=&\underset{\pi\in\Mm_+(\Xx\times\Xx)}{\min} \Bigg\{ \int_{\Xx\times\Xx} d(\V x,\V y) \mathrm{d}\pi(\V x,\V y)\\ \notag
	&+ \la \mnorm{\mu - \pushx{\pi}} + \la \mnorm{\nu - \pushy{\pi}}\Bigg\},
\end{align}
where $\pi\in\Mm_+(\Xx\times\Xx)$ is a nonnegative Radon measure over $[0,1]^\dim \times [0,1]^\dim$ that specifies the transport plan between the marginals $\pushx{\pi}=\int_\Xx \mathrm{d}\pi(\cdot,\V y)$ and $\pushy{\pi}=\int_\Xx \mathrm{d}\pi(\V x,\cdot)$ of $\pi$ (which can be made arbitrary close to $\mu$ and $\nu$, respectively,  by setting $\lambda\to+\infty$).
\end{proposition}

\begin{figure}[t]
    \centering
    \def\lam{.8}
    \def\a{.85}
    \def\b{.55}
    \begin{tikzpicture}[scale=0.5]
        \node[anchor = east] at (0,2.9) {a)};
        \draw[-latex](-.1,0) -- (4,0);
        \draw[-latex](0,-.1) -- (0,3);
        \node (x1) [black, opacity = \a] at (2,1) {$\bullet$};
        \node[anchor=west] at (x1) {\small $\V{x}_1$}; 
        \draw[thin,dotted] (x1) circle (2*\lam);
        \draw[<->] (x1) -- +(-1.3*\lam,-1.3*\lam) node [midway,above,xshift=-5pt] {\small $2\lambda$};
        \node (y1) [black, opacity = \b] at (2.3,1.6) {$\bullet$};
        \node[anchor=west] at (y1) {\small $\V{y}_1$}; 
    \end{tikzpicture}
    \begin{tikzpicture}[scale=0.5]
        \draw[-latex] (-.1,0) -- (4,0) node [anchor=west] {$\|\V{x}_1-\V{y}_1\|_2$};
        \draw[-latex] (0,-.1) -- (0,\a*\lam+\b*\lam+.5) node (title) [anchor = south]{\small $\KR{\mu}{\nu}$};
        \node [anchor=east] at (title.west) {b)};
        \draw (0, \a*\lam-\b*\lam) -- (2*\lam,\a*\lam+\b*\lam) -- (3.8,\a*\lam+\b*\lam);
        \node[anchor=north] at (0,-.1) {\small $0$};
        \draw (2*\lam,0) -- +(0,-.1) node [anchor=north] {\small $2\lambda$};
        \draw (0,\a*\lam-\b*\lam) -- +(-.1,0) node [anchor=east] {\small $(a_1-b_1)\lambda$};
        \draw (0,\a*\lam+\b*\lam) -- +(-.1,0) node [anchor=east] {\small $(a_1+b_1)\lambda$};
    \end{tikzpicture}
    \caption{\label{fig:flatmetric_twoDiracs}a) Example of two discrete measures in $\mathcal{X} = [0,1]^2$, $\mu = a_1 \dirac{\V{x}_1}$ and $\nu = b_1 \dirac{\V{y}_1}$, where $a_1$ and $b_1$ are represented by the opacity of the $\bullet$ marks. b) Dependence of the metric $\KR{\mu}{\nu}$ on $\|\V{x}_1-\V{y}_1\|_2$ from a) for fixed values $a_1>b_1$, growing linearly with $\|\V{x}_1-\V{y}_1\|_2$ and saturating at $\|\V{x}_1-\V{y}_1\|_2=2\lambda$.\label{fig:Flat}}
\end{figure}
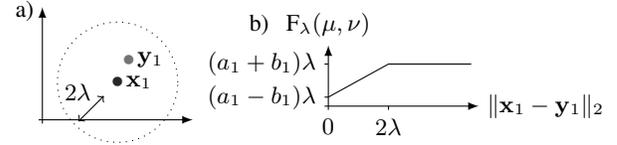

The first term in the minimization problem~\eqref{eq:dual-primal-continuous} penalizes the cost of transporting $\pushx{\pi}$ to $\pushy{\pi}$ (or vice versa). This is, in fact, the same cost function as in the $1$-Wasserstein distance, one of the classical optimal-transport problems. Optimal-transport metrics quantify how different two measures are by assessing the cost of transforming (in other words, transporting) one measure onto the other.
Unlike in standard optimal transport, the marginals $\pushx{\pi}$ to $\pushy{\pi}$ need not be equal to the measures of interest $\mu$ and $\nu$. Instead, the constraints are relaxed using the second and third discrepancy terms in \eqref{eq:dual-primal-continuous} which involve the total-variation norm. This relaxation allows for the creation and destruction of mass before transport and, therefore, for an optimal transport between measures with different total mass. This key feature is essential for SMLM, as it accounts for the errors both of localization (by the cost of transport) and of detection (by the cost of creation or destruction of mass). Their balance is controlled by the physically interpretable parameter 
$\lambda > 0~[\nano\metre]$, 
as illustrated in Figure~\ref{fig:flatmetric_twoDiracs}. 
When the two Dirac masses are at the same position, the cost is proportional to the difference of weights. Then, it grows linearly with $\norm{\V{x}_1-\V{y}_1}_2$ as the Dirac mass $b_1\dirac{\V{y}_1}$ is transported to the position $\V{x}_1$. This keeps happening until $\norm{\V{x}_1-\V{y}_1}_2\geq 2\lambda$, where the masses are no longer moved and the cost results from the pure creation and destruction of mass.

It is also important to note that Flat Metric is homogeneous to nanometers so that it can be physically associated to a specific scale (in nanometers for the SMLM problem).
Hence, when the number of locations is estimated correctly, Flat Metric represents the mean error in terms of localization, similar to the RMSE (see Figure~\ref{fig:surface}).
When $\lambda\to+\infty$ and $\mu$ and $\nu$ have the same mass, we recover the $1$-Wasserstein distance ($\|\cdot\|_{W_1}$). Finally, when $\lambda\to0$, we recover the total-variation norm.
Consequently, Flat Metric is an interpolating distance between $\|\cdot\|_{W_1}$ and $\mnorm{\cdot}$.

\subsection{How to Compute Flat Metric}

The ground-truth data can be represented as the discrete Radon measure
\begin{align}\label{eq:measure-gt}
	\mu = \sum_{n=1}^N a_n\dirac{\V{x}_n}\in\Radon\qwithq a_n>0, \V{x}_n\in\Xx,
\end{align}
which contains the locations of the fluorescent emitters in a frame. The reconstructed locations given by any software can also be represented as the discrete Radon measure
\begin{align}\label{eq:measure-est}
	\nu = \sum_{m=1}^M b_m\dirac{\V{y}_m}\in\Radon\qwithq b_m>0, \V{y}_m\in\Xx.
\end{align}
In this discrete setting, we simplify the computation of Flat Metric $\KR{\mu}{\nu}$ in \eqref{eq:def-flatmetric-general} as detailed in Proposition~\ref{prop:discrete-problem}.

\begin{proposition}\label{prop:discrete-problem}
When $\mu$ and $\nu$ are discrete Radon measures, one can compute \eqref{eq:def-flatmetric-general} as
\begin{align}\label{eq:kr-dual-simplified}
	\KR{\mu}{\nu}=-\min_{\V f \in\RR^{N+M}\text{\emph{ s.t. }} \Op L \V f \in \Bb\times\Cc} \ \dotp{\V f}{-\V c},
\end{align}
where $\Op L:\RR^{N+M}\rightarrow \RR^{N\times M}\times\RR^{N+M}$ is defined by
\begin{align}\label{eq:operatorL}
	\Op L \V f = \left(\begin{pmatrix} \V f_n - \V f_{N+m}\end{pmatrix}_{1\leq n\leq N, 1\leq m\leq M}, \V f\right),
\end{align}
where $\V c = (\V a, -\V b)\in\RR^{N+M}$, and $\Bb$ and $\Cc$ are hyper-rectangles such that $\Bb=\{\V G\in\RR^{N\times M};\ \V |G_{n,m}| \leq d(\V{x}_{n},\V{y}_{m})\}$ and $\Cc=\{\V f\in\RR^{N+M}:\ \forall k\in\{1,2,\ldots,N+M\},|f_k|\leq \lambda\}$.

In fact,~\eqref{eq:kr-dual-simplified} holds because the dual problem of that minimization is exactly the unbalanced optimal transport problem~\eqref{eq:dual-primal-continuous} for discrete measures, and strong duality holds.
\end{proposition}

Therefore, to compute $\KR{\mu}{\nu}$, one simply needs to solve the minimization problem given in \eqref{eq:kr-dual-simplified}, which is a finite-dimensional linear program.
This problem is then solved using any standard linear programming toolbox.

Note that if one considers only the first part of the operator $\Op L$ in~\eqref{eq:operatorL} then~\eqref{eq:kr-dual-simplified} is exactly the dual problem of the $1$-Wasserstein optimal transport problem, see~\cite[ch.~6]{Peyre2019Computational}. The second part accounts for the relaxation allowing creation and destruction of mass, as explained above.



\section{Numerical Experiments} 
\label{sec:exp}
\DTLnewdbonloadfalse
\DTLnewdb{dets}
\DTLnewdb{GT}
\DTLnewdb{transport}

In this section, we first propose an example to illustrate the behavior of Flat Metric. Then, we detail how we generated Figure~\ref{fig:surface}, which confirms that Flat Metric has a behavior similar to that of the efficiency~\cite{Sage2019} and that it provides a continuum between detection and localization errors. Finally, we report Flat Metric as obtained by 31 participants of the SMLM 2016 Challenge on the $2$D dataset and compare it with their efficiency and RMSMDs.
Note that, in all our experiments, the weights of the ground-truth are uniform and the obtained scale is applied for the reconstruction, with $a_n=b_m=1/N$. 
We use the normalizing scaling of the ground-truth for the reconstruction as it provides a coherent way to compare different software which do not detect the same number of point sources.


\subsection{Interpretation of Flat Metric}

\DeclareRobustCommand\GT{\tikz \draw[fill=red,opacity=0.5,draw opacity=0] (0,0) circle (.1);}
\DeclareRobustCommand\DET{\tikz \draw[fill=green,opacity=0.5,draw opacity=0] (0,0) circle (.1);}

\begin{figure}
     \centering
     \input{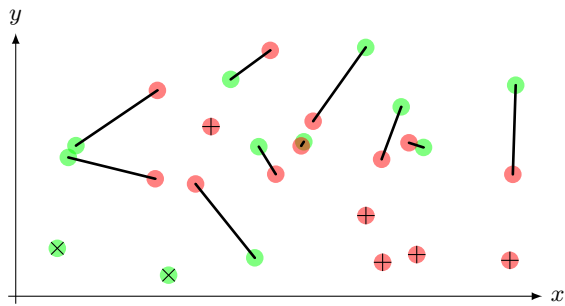}
     \caption{Illustration of how Flat Metric (here equal to $0.125$) is computed. 
     When an estimated location (\DET) is linked by a line to a ground-truth location (\GT), the cost
	 ($d(\V x, \V y)/N$) comes from moving the former to the latter.
	 The presence of a cross ($\times$) means that the point has been destroyed, at the cost of $\lambda/N$. A ground-truth location with a plus sign (+)
	 means a mass has been created at this position to match it, also at the cost of $\lambda/N$.}
     \label{fig:toy_example}
\end{figure}

We show in Figure~\ref{fig:toy_example} an example of a ground-truth dataset (\GT), and its reconstructed dataset (\DET), and how Flat Metric accounts for the difference between these two measures. Ground-truth locations were chosen randomly in the rectangle $[0,1]\times[0,0.5]$ with weights $a_n=b_m=1/N$ with $N=15$. 
Here, $\lambda=0.1$, which constrains the maximal transport distance between two isolated point sources to~$0.2$ (see Figure~\ref{fig:Flat}).
Our interpretation of Flat Metric comes from its link with unbalanced optimal transport (see Proposition~\ref{prop:uot}). As a metric, it is symmetric. Therefore, we arbitrarily
choose to interpret it as the cost of transporting the estimation towards the ground truth (GT). As a result, we have the following behaviors.
\begin{itemize}
\item \emph{Transport}: A Dirac mass $\dirac{\V y}/N$ of the reconstruction is moved towards one in the ground-truth data $\dirac{\V x}/N$. The cost of this transport is $d(\V x, \V y)/N$.
\item \emph{Destruction of mass}: A Dirac mass $\dirac{\V y}/N$ of the reconstruction is destroyed because there is no corresponding ground-truth location nearby. The cost of this destruction of mass is $\lambda \mnorm{\dirac{\V y}/N} = \lambda/N$.
\item \emph{Creation of mass}: A Dirac mass $\dirac{\V y}/N$ is created at a position $\V y$ to match a ground-truth location when there is no corresponding Dirac mass in the reconstruction. This cost is $\lambda/N$.
\end{itemize}

Note that we have only these three alternatives because of our choice of weights. To have more complex phenomena such as simultaneous transport, creation, and destruction of mass, discrete measures with 
Dirac masses of different weights should be used. This could certainly be of interest to the evaluation of other point-source localization problems.

\subsection{Synthetic Experiments on Flat Metric and Efficiency} 
    
    We show in Figure~\ref{fig:surface} how Flat Metric, just as efficiency, interpolates between detection and localization metrics. To its benefit,
    Flat Metric has strong foundations in the theory briefly presented in
    Section~\ref{sec:math}, by contrast with the efficiency measure which is based on empirical results.
    Consistently, Flat Metric is also well-defined for $0\%$ recall, thus being
    a robuster tool for any use-case.

    In order to exhibit this link in conditions relevant to SMLM, we chose to
    focus on recall as a detection metric. Indeed, recall is typically
    the most relevant factor to characterize detection in SMLM, as most leading
    algorithms achieve very high precision~\cite{Sage2019}. We modeled this 
    situation by randomly sampling $100$ ground-truth locations uniformly in
    a square of $(6.4 \times 6.4)~\mu\metre$, and simply removing the
    corresponding percentage of locations to initialize the set of recovered
    locations.
    
    For the joint evaluation of detection and localization effects, we
    modeled localization errors in detected locations as independent and identically distributed 
    uniform vectors in disks of radius up to 
    $250~\nano\metre$.  
    
    The results in Figure~\ref{fig:surface} were generated by averaging $50$ 
    randomized trials for each combination of radius and recall, using 
    $\lambda = 125~\nano\metre$. 
    Finally, the expectations of Flat Metric and efficiency are shown on the planes with $100\%$
    recall and vanishing perturbation radius, respectively. They are related to 
    the expectations of $\mathrm{J}$ and $\mathrm{RMSE}$ in those cases,
    where $c=2/3\sqrt{2}$. 

\subsection{Application to the 2016 Challenge}
\begin{figure}
	\centering

 \begin{tikzpicture}
    	\newcommand{\marksize}{0.75pt}
     	\begin{axis}[name = wtf1,
    		width = {0.4*\textwidth},
    		height = 2.65cm,
    		ticks = none, draw=none, hide axis,
		legend columns=2,
    		legend style={
			at={(0.5,0.38)},
    			anchor=base,
    			column sep=0.5em,
    			legend style={font=\scriptsize,},
  		},
 	 ]
    		\addlegendimage{only marks,mark=*,red,mark size=\marksize,opacity=0.5,
			draw opacity=0,};
    		\addlegendentry{ER1.N3.LD};
    		\addlegendimage{only marks,mark=*,blue,mark size=\marksize,opacity=0.5,
			draw opacity=0,};
    		\addlegendentry{ER2.N3.HD};
    		\addlegendimage{only marks,mark=*,green!50!black,mark size=\marksize,opacity=0.5,
			draw opacity=0,};
    		\addlegendentry{MT3.N2.LD}
    		\addlegendimage{only marks,mark=*,black,mark size=\marksize,opacity=0.5,
			draw opacity=0,};
    		\addlegendentry{MT4.N2.HD};
		\addplot [draw=none] {x};
    \end{axis}
    \begin{groupplot}[
    	group style={
		group name=wtf, 
		group size={2 by 2},
    		vertical sep=5pt,
		horizontal sep=5pt,
	},
	width=0.29*\textwidth,
	height=0.29*\textwidth,
	x label style={at={(axis cs:75,2000)},anchor = north},
	y label style={at={(axis cs:-30,0)},anchor = base},
	label style={font=\scriptsize},
	xmin = -10, xmax=70,
    	ymin = 0, ymax=1800,
	ticklabel style = {font=\scriptsize},
    ]
    \nextgroupplot[
		y dir=reverse,
		xticklabels={},
    		at = (wtf1.south),
    		anchor = north east,
    	]
	    \addplot[
			only marks,
    			mark=*,
    			color=red,
    			mark size=\marksize,
			opacity=0.5,
			draw opacity=0,
		] table [x=efficiency, y=rmsmd, col sep=comma]{figs/challenges/ER1.N3.LD_RMSMD.csv};
    
    \nextgroupplot[y dir=reverse,xticklabels={,,},
    yticklabels={,,},
    ]
    \addplot[
			only marks,
    			mark=*,
    			color=blue,
    			mark size=\marksize,
			opacity=0.5,
			draw opacity=0,
		] table [x=efficiency, y=rmsmd, col sep=comma] {figs/challenges/ER2.N3.HD_RMSMD.csv};
    
    \nextgroupplot[y dir=reverse, ylabel={RMSMD},
    xlabel={Efficiency},
    ]
    \addplot[
			only marks,
    			mark=*,
    			color=green!50!black,
    			mark size=\marksize,
			opacity=0.5,
			draw opacity=0,
		] table [x=efficiency, y=rmsmd, col sep=comma] {figs/challenges/MT3.N2.LD_RMSMD.csv};
    
    \nextgroupplot[y dir=reverse, yticklabels={,,}, 
    ]
    \addplot[
			only marks,
    			mark=*,
    			color=black,
    			mark size=\marksize,
			opacity=0.5,
			draw opacity=0,
		] table [x=efficiency, y=rmsmd, col sep=comma]  {figs/challenges/MT4.N2.HD_RMSMD.csv};
    \end{groupplot}%
    \end{tikzpicture}    
	\caption{\label{fig:challengeRMSMD} Low degree of correlation between efficiency and the RMSMD on four datasets from the SMLM 2016 Challenge.}
\end{figure}
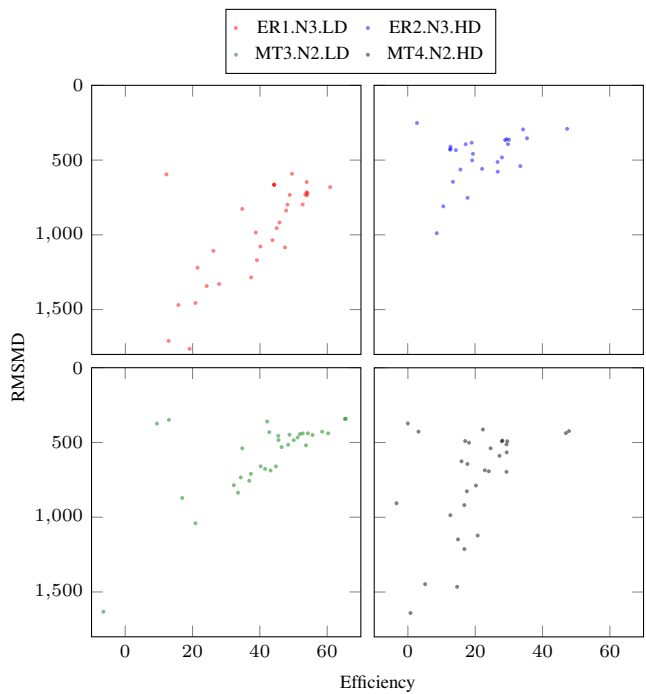

We compare efficiency, a thoroughly validated empirical metric for SMLM, to both Flat Metric and RMSMD, on four 2D datasets from the SMLM 2016 Challenge\footnote{\hyperlink{http://bigwww.epfl.ch/smlm/challenge2016/index.html?p=results}{http://bigwww.epfl.ch/smlm/challenge2016/index.html?p=results}}.
As shown in Figures.~\ref{fig:surface}~and~\ref{fig:challengeRMSMD}, Flat Metric is strongly correlated with efficiency, while RMSMD is not. 
Indeed, one only observes few outliers on the efficiency vs. Flat Metric comparison, mainly for reconstruction methods that
work rather poorly on the datasets~MT3 and MT4.


%
\section{Conclusion}

We propose Flat Metric to quantitate SMLM reconstruction errors when ground-truth data are available. Here, we present and exemplify the strong links between Flat Metric and unbalanced optimal transport problems, which underpin this robust metric. We also provide exhaustive evidence that Flat Metric is conceptually similar to efficiency, a very well established empirical metric designed in the organization of 
the SMLM 2016 Challenge. Consequently, we provide a robust and practical metric for SMLM evaluation.
We have also exemplified and explained how Flat Metric works internally, providing intuition on how 
this optimal assessment is obtained. Further, we have emphasized the interpretability of Flat Metric,
which can be read as an equivalent localization accuracy.

\section{Compliance with Ethical Standards}
This article does not contain any studies involving human participants or animals performed by any of the authors.

\section{Acknowledgments}

The authors would like to thank Gabriel Peyr\'e and Vincent Duval who first suggested to use Flat Metric for SMLM and encouraged the authors to pursue this direction.

We acknowledge access to the facilities and expertise of the CIBM Center for Biomedical Imaging, a Swiss research center of excellence founded and supported by Lausanne University Hospital (CHUV), University of Lausanne (UNIL), Ecole polytechnique fédérale de Lausanne (EPFL), University of Geneva (UNIGE) and Geneva University Hospitals (HUG).

There is no potential conflict of interest.

\bibliographystyle{IEEEtran}
\bibliography{references}
\end{document}